\documentclass[aps,prb,twocolumn,showpacs,superscriptaddress]{revtex4}
\usepackage{graphicx}
\begin{document}

\title{Anisotropic Neutron Spin Resonance in the Fe-based 
BaFe$_{1.9}$Ni$_{0.1}$As$_2$ Superconductor}

\author{O. J. Lipscombe}
\email{lipscombe@utk.edu}
\affiliation{Department of Physics and Astronomy, The University of Tennessee, Knoxville,
Tennessee 37996-1200, USA }
\author{Leland W. Harriger}
\affiliation{Department of Physics and Astronomy, The University of Tennessee, Knoxville,
Tennessee 37996-1200, USA }
\author{P. G. Freeman}
\affiliation{Institut Laue-Langevin, 6, rue Jules Horowitz, BP156-38042 Grenoble Cedex 9, France}
\author{M. Enderle}
\affiliation{Institut Laue-Langevin, 6, rue Jules Horowitz, BP156-38042 Grenoble Cedex 9, France}
\author{Chenglin Zhang}
\affiliation{Department of Physics and Astronomy, The University of Tennessee, Knoxville,
Tennessee 37996-1200, USA }
\author{Miaoying Wang}
\affiliation{Department of Physics and Astronomy, The University of Tennessee, Knoxville,
Tennessee 37996-1200, USA }
\author{Takeshi Egami}
\affiliation{Department of Physics and Astronomy, The University of Tennessee, Knoxville,
Tennessee 37996-1200, USA }
\affiliation{Department of Materials Science and Engineering, The University of Tennessee, Knoxville,
Tennessee 37996-1200, USA }
\affiliation{Oak Ridge National Laboratory, Oak Ridge, Tennessee 37831}
\author{Jiangping Hu}
\affiliation{
Department of Physics, Purdue University, West Lafayette, IN 47907, USA}
\affiliation{Institute of Physics, Chinese Academy of Sciences, Beijing 100190, China
}
\author{Tao Xiang}
\affiliation{Institute of Physics, Chinese Academy of Sciences, Beijing 100190, China
}
\affiliation{Institute of Theoretical Physics, Chinese Academy of Sciences, P.O. Box 2735, Beijing 100190, China}
\author{M. R. Norman}
\affiliation{
Materials Science Division, Argonne National Laboratory, Argonne, Illinois 60439, USA
}
\author{Pengcheng Dai}
\email{daip@ornl.gov}
\affiliation{Department of Physics and Astronomy, The University of Tennessee, Knoxville,
Tennessee 37996-1200, USA }
\affiliation{Oak Ridge National Laboratory, Oak Ridge, Tennessee 37831}
\affiliation{Institute of Physics, Chinese Academy of Sciences, Beijing 100190, China
}
\begin{abstract}
We use polarized inelastic neutron scattering to show that the
neutron spin resonance below $T_c$ in superconducting
BaFe$_{1.9}$Ni$_{0.1}$As$_2$ ($T_c=20$ K) is purely magnetic in origin.
Our analysis further reveals that the resonance peak near
7~meV only occurs for the planar response.  This challenges the 
common perception that the spin resonance in the pnictides is an isotropic
triplet excited state of the singlet Cooper pairs, as our results imply that only the
$S_{001}=\pm1$ components
of the triplet are involved.
\end{abstract}

\pacs{74.70.Xa, 78.70.Nx, 75.30.Gw}

\maketitle

\section{Introduction}
The neutron spin resonance is a collective magnetic excitation appearing
in the superconducting state of high-transition temperature (high-$T_c$) copper oxide superconductors \cite{mignod,mook,fong,dai}.
 Since its initial discovery in optimal hole-doped YBa$_2$Cu$_3$O$_{6+x}$ \cite{mignod,mook,fong,dai},
the resonance has been found in electron-doped cuprates \cite{wilson}, heavy fermion \cite{metoki,stock}, and iron arsenide superconductors
\cite{christianson,lumsden,chi,slli,inosov,mywang}. Below the superconducting transition temperature $T_c$, the intensity of the resonance increases like the superconducting  order parameter and its energy scales with $T_c$ \cite{wilson}.
Although the resonance appears to be a ubiquitous property of unconventional superconductors \cite{mignod,mook,fong,dai,wilson,metoki,stock,christianson,lumsden,chi,slli,inosov}, its microscopic origin and relationship
with superconductivity are still debated \cite{eschrig}.  In all these materials, the resonance occurs at the antiferromagnetic (AF) wavevector $\mathbf{Q}$
of the parent compound.  It is thought to be a triplet excitation of the singlet 
Cooper pairs \cite{eschrig,so5}, 
implying a superconducting order parameter that 
satisfies  $\Delta_{k+Q}=-\Delta_k$.  In the iron arsenide superconductors, this condition is satisfied
by an order parameter whose sign reverses between the electron and hole 
pockets \cite{mazin,korshunov,maier1,maier2,seo2}.
If this picture is correct,
one would expect that the triplet would be degenerate, and thus directionally isotropic in space.
For the optimal hole-doped high-$T_c$ cuprate superconductor YBa$_2$Cu$_3$O$_{6+x}$,
polarized inelastic neutron scattering experiments suggest that this is indeed the case \cite{mook,fong}.

We report polarized inelastic neutron scattering
results for the optimal electron-doped iron arsenide superconductor
BaFe$_{1.9}$Ni$_{0.1}$As$_2$ ($T_c=20$~K) \cite{chi,slli}.  We find
that the resonance previously observed around 7 meV at the AF
wavevector $\mathbf{Q}=(0.5,0.5,1)$ (reciprocal lattice units, `r.l.u') is
entirely magnetic, but displays strong spin-space anisotropy, with a peaked response
near the resonance energy
occurring only for the planar response.
This is different from the momentum space anisotropy, where the spin correlation length might be
different along different crystallographic directions \cite{lester10,hfli,diallo}.
Our results
indicate a strong spin-orbital/lattice coupling in iron arsenide
superconductors (quite different from the cuprates), and
are a challenge to the common assumption that the resonance
represents an isotropic singlet-to-triplet excitation.

\section{Experimental Details}
\subsection{Sample}
We chose the iron arsenide superconductor BaFe$_{1.9}$Ni$_{0.1}$As$_2$ because
this material has no static AF order [Fig.~1(b)], exhibits a well-defined neutron spin resonance near 7 meV at $\mathbf{Q}=(0.5,0.5,1)$ above a clear spin gap, and
is available in large, homogeneous single crystals \cite{chi,slli}.
We co-aligned $\sim$5 grams of single crystals (with mosaic of 3$^{\circ}$ full-width-half-maximum) in the $(H,H,L)$ scattering plane \cite{chi,slli}, where the wavevector $\mathbf{Q}$ is indexed
$\mathbf{Q}=H \mathbf{a^{\star}} + K \mathbf{b^{\star}} + L \mathbf{c^{\star}}$ with $\mathbf{a^{\star}}=\hat{\mathbf{a}}\ 2\pi/a$, etc., $a=b=3.93$ \AA\
and $c=12.77$ \AA\ [Fig.~1(a)].  In this tetragonal notation, the AF order and resonance occur at $\mathbf{Q}=(0.5,0.5,L)$ with $L=\pm 1,\pm 3,\cdots$ \cite{lumsden,chi,slli}.

\begin{figure}
\begin{center}
\includegraphics[width=0.8\linewidth]{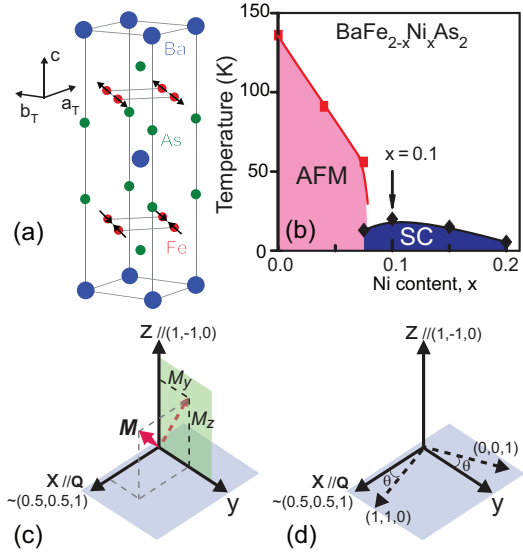}
\end{center}\caption{(color online) (a) Crystal structure of BaFe$_2$As$_2$. (b) Magnetic and superconducting phase diagram of BaFe$_{2-x}$Ni$_{x}$As$_2$
with the present composition highlighted with an arrow \cite{mywang}. (c) Schematic showing a fluctuating atomic magnetic moment vector $\mathbf{m}$, and the components $m_y$ and $m_z$ which are probed. Neutron scattering intensity is related to the square of the fluctuating moment components in the $y$ and $z$ direction, $M_y=\langle m_y^2 \rangle$ and $M_z=\langle m_z^2\rangle$ respectively (which are both defined perpendicular to $x$, where $x$ is parallel to the wavevector $\mathbf{Q}$). (d) Schematic showing crystallographic in-plane $(1,1,0)$, $(1,-1,0)$ and out of plane $(0,0,1)$ directions compared with the $x$, $y$, $z$ directions defined above.}
\label{Fig:fig1}
\end{figure}

\subsection{Polarized Neutron Analysis}
We carried out polarized inelastic neutron scattering experiments using the Cryopad capability of the
IN20 triple-axis spectrometer at the Institut Laue-Langevin, Grenoble, France.
Neutron polarization analysis is the only way to conclusively separate the magnetic signal from lattice effects, and to determine the spatial anisotropy of the magnetic excitations. 

In principle, polarization analysis can be used to completely separate magnetic (e.g.~spin fluctuation) and nuclear (e.g.~phonon) scattering because the spin of the neutron is always flipped in a magnetic interaction where the neutron polarization is parallel to the wavevector transfer $\mathbf{Q}$  \cite{moon}.
For convenience, we define the neutron polarization directions along $\mathbf{Q}$ as
$x$, perpendicular to $\mathbf{Q}$ but in the scattering plane as $y$, and perpendicular to $\mathbf{Q}$
and the scattering plane as $z$, respectively [Fig.~1(c)]. 
At a specific wavevector and energy, we
measured the six cross-sections which correspond to the
three incoming neutron polarization directions $x$, $y$, $z$,
with the outgoing neutron polarization either parallel to
the incoming (neutron non-spin flip or NSF) or anti-parallel (neutron spin-flip or SF). The measured neutron cross-sections
are then accordingly written as $\sigma_{\alpha}^{\rm NSF}$ and $\sigma_{\alpha}^{\rm SF}$ where $\alpha=x,y,z$  \cite{moon}. With the Cryopad setup, these cross-sections can 
be measured with the sample in a strictly zero magnetic field ($<10$ mG), thus avoiding errors due to flux inclusion or field 
expulsion in the superconducting phase of the sample.

We define the magnetic intensity of excitations with fluctuating magnetic moments pointing parallel to the $(1,1,0)$ (in-plane) direction as $M(110)$, and the intensity of fluctuating moments pointing out of plane as $M(001)$. Our experiment probes $M_y$ and $M_z$, the magnetic intensity of excitations with the moment parallel to $y$ and $z$ respectively [see Fig.~\ref{Fig:fig1}(c)]. Due to tetragonal symmetry $M(110)=M(1\bar{1}0)\equiv M_z$, and $M(001)$ can be found from $M_y$ using $M_y=M(110)\sin^2\theta+M(001)\cos^2\theta$, where $\theta$ is the angle between wavevectors
$(1,1,0)$ and $\mathbf{Q}$ [see Fig.~\ref{Fig:fig1}(d)].

The measured cross-section can be written
\begin{figure}
\begin{center}
\includegraphics[width=0.9\linewidth]{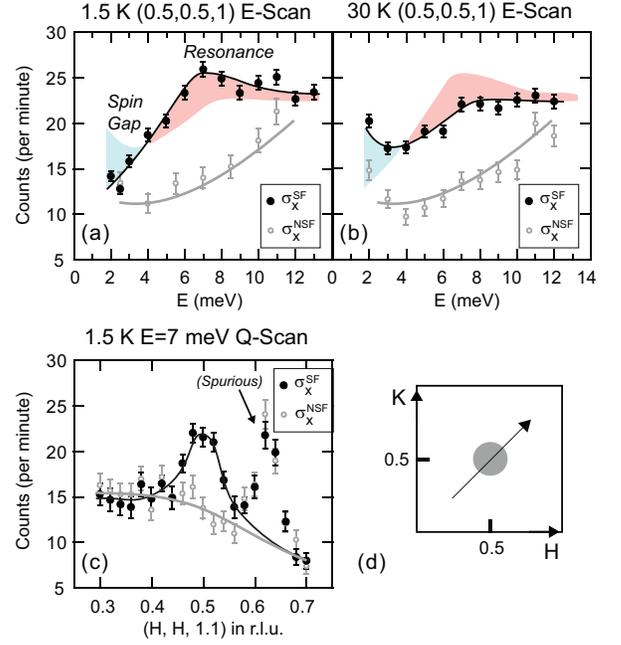}
\end{center}\caption{(color online) Energy scans at $\mathbf{Q}=(0.5,0.5,1)$, showing $\sigma_{x}^{\rm SF}$ (magnetic) and $\sigma_{x}^{\rm NSF}$ (nuclear) scattering for (a) 1.5~K and (b) 30~K. (c) $(H,H,1.1)$ $Q$-Scan through the resonance position, showing $\sigma_{x}^{\rm SF}$ and $\sigma_{x}^{\rm NSF}$ measured at a constant energy 7~meV. The narrow peak at $(0.625,0.625)$ is temperature independent spurious scattering. (d) Trajectory in reciprocal space of the $(H,H,1.1)$ scan. Solid lines are guides to the eye for all plots except where otherwise stated.}
\label{Fig:fig2}
\end{figure}

\begin{equation}
\label{Eq:XS}
\left( \begin{array}{c}
\sigma_x^{\rm SF}-b_1\\
\sigma_y^{\rm SF}-b_1\\
\sigma_z^{\rm SF}-b_1\\
\sigma_x^{\rm NSF}-b_2\\
\sigma_y^{\rm NSF}-b_2\\
\sigma_z^{\rm NSF}-b_2
\end{array} \right)=
\frac{1}{R+1}\left( \begin{array}{ccccc}
R & R & 1 \\
1 & R & 1 \\
R & 1 & 1 \\
1 & 1 & R \\
R & 1 & R \\
1 & R & R
\end{array} \right)
\left( \begin{array}{c}
M_y\\
M(110)\\
N\\
\end{array} \right),
\end{equation}
with a nuclear scattering strength $N$ (containing both phonon and inelastic incoherent nuclear scattering), and $b_1$ and $b_2$ account for instrumental background (and nuclear-spin incoherent scattering). $R$ specifies the quality of the neutron beam polarization (so that leakage between SF and NSF channels caused by imperfect polarization are taken into account). In our setup we measured $R$ by the leakage of nuclear Bragg peaks into the (magnetic) SF channel $R=\sigma_{\rm Bragg}^{\rm NSF} / \sigma_{\rm Bragg}^{\rm SF}\approx$ 15, independent of neutron polarization direction. To extract $M_y$ and $M(110)$ from the raw data we can use (from Eq.~1)
\begin{eqnarray}
\sigma_x^{\rm SF}-\sigma_y^{\rm SF}=\sigma_y^{\rm NSF}-\sigma_x^{\rm NSF}=c M_y, \nonumber \\
\sigma_x^{\rm SF}-\sigma_z^{\rm SF}=\sigma_z^{\rm NSF}-\sigma_x^{\rm NSF}=c M(110),  
\end{eqnarray}
where $c=(R-1)/(R+1)$. With Eq.~2, we can estimate the energy and wavevector dependence of $M(110)$ or $M_y$ from a weighted average of the pair of values calculated using the SF and the NSF data.

\begin{figure}
\begin{center}
\includegraphics[width=0.9\linewidth]{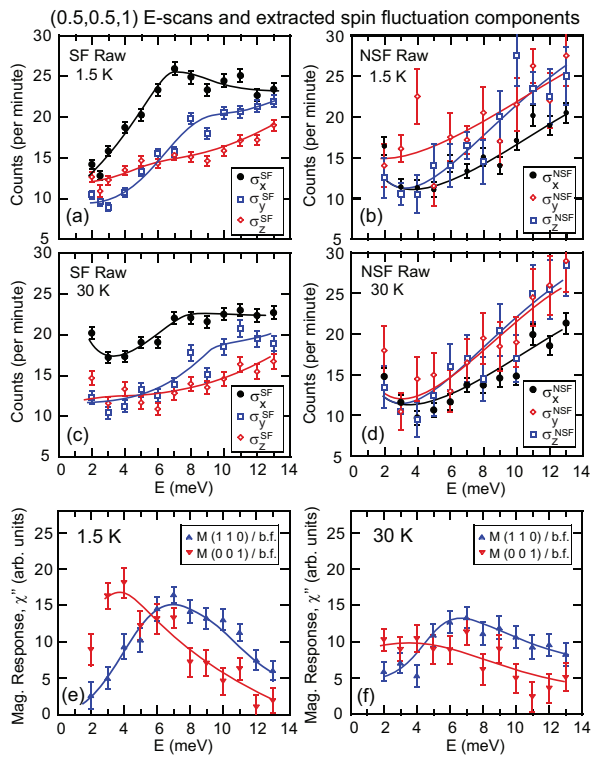}
\end{center}\caption{(color online) Energy scans at $\mathbf{Q}=(0.5,0.5,1)$. Raw (a) 1.5~K $\sigma_{x,y,z}^{\rm SF}$ and (b) 1.5~K $\sigma_{x,y,z}^{\rm NSF}$ cross-section data. Clear anisotropy is evident because $\sigma_{y}^{\rm SF}\neq \sigma_{z}^{\rm SF}$ and $\sigma_{y}^{\rm NSF}\neq \sigma_{z}^{\rm NSF}$. (c,d) Raw data taken at 30 K.
(e)-(f) In-plane and out of plane magnetic response [the extracted $M(110)$ and $M(001)$ using raw data in Eq.~2, divided by the Bose factor (`b.f.')] at 1.5~K and 30~K respectively. Data in (e) and (f) are also corrected for second-order monitor over-counting.
}
\label{Fig:fig3}
\end{figure}

\section{Results}
Figs.~2(a)-(b) show $\sigma_{x}^{\rm SF}$ (primarily magnetic) and $\sigma_{x}^{\rm NSF}$ (primarily nuclear) energy cuts at $(0.5,0.5,1)$ taken at temperatures of 1.5~K ($\ll T_c$) and 30~K ($>T_c$), respectively. As the temperature decreases, it is clear that the nuclear scattering ($\sigma_{x}^{\rm NSF}$) changes very little with temperature, while the magnetic scattering ($\sigma_{x}^{\rm SF}$) around 7~meV is enhanced, and below $\sim$3~meV becomes gapped \cite{chi,slli}. These data unambiguously demonstrate that the resonance is purely magnetic without any lattice contribution.
Fig.~2(c) shows a $T$=1.5~K $Q$ cut along the $(H,H,1.1)$ trajectory at 7~meV [Fig.~2(d)].  Consistent with unpolarized
measurements \cite{chi,slli}, the data prove that the resonance is magnetic scattering centered at $(H,K)=(0.5,0.5)$.

Having established the magnetic nature of the resonance, we now probe the anisotropy of the spin fluctuation spectrum by measuring $\sigma_{x,y,z}^{\rm SF}$ and $\sigma_{x,y,z}^{\rm NSF}$ and using Eq.~2 to calculate $M(110)$ and $M(001)$.
$\sigma_{y}^{\rm SF}$ exclusively probes the in-plane
spin fluctuations $M(110)$
and $\sigma_{z}^{\rm SF}$ gives the intensity of moments fluctuating along $M_y\sim M(001)$.
Finally, $\sigma_{x}^{\rm SF}$ is the magnetic part of the cross-section observed in unpolarized measurements and provides the sum of the magnetic scattering, in this case $M_y+M(110)$. For isotropic
paramagnetic spin fluctuations, one expects $M_y=M(110)$ and this appears to be the case
for the resonance in optimal doped YBa$_2$Cu$_3$O$_{6+x}$ \cite{mook,fong}.

Figs.~3(a)-(d) show all six scattering cross-sections
$\sigma_{x,y,z}^{\rm SF}$ and $\sigma_{x,y,z}^{\rm NSF}$ raw data taken
at $\mathbf{Q}=(0.5,0.5,1)$ below and above $T_c$. While the resonance at 7 meV is clearly seen in the
$\sigma_{x}^{\rm SF}$ data at 1.5 K [Fig.~3(a)], a comparison of $\sigma_{y}^{\rm SF}$
and $\sigma_{z}^{\rm SF}$ shows that the former has a peak
while the latter is featureless near the resonance energy. Since $\sigma_{z}^{\rm SF}\sim M(001)$ and $\sigma_{y}^{\rm SF}\propto M(110)$,
these data immediately suggest anisotropic spin fluctuations near the resonance.
By using all
six scattering cross-sections in Figs.~3(a) and 3(b), we extract the energy dependence of
$M(110)$ and $M(001)$ magnetic scattering, and subsequently convert the extracted data to a magnetic
response, $\chi^{\prime\prime}_{110}$ and $\chi^{\prime\prime}_{001}$, [Fig. 3(e)] by dividing out the Bose population factor [also, we can instead extract $M(110)$ and $M(001)$ from only the three SF cross-sections, in which case we get quantitatively very similar results]. 
It is clear that the in-plane response ($\chi^{\prime\prime}_{110}$) resembles a peak centered at around 7~meV, whilst the out of plane $\chi^{\prime\prime}_{001}$ has a much lower energy scale.

Figs.~3(c) and 3(d) show
$\sigma_{x,y,z}^{\rm SF}$ and $\sigma_{x,y,z}^{\rm NSF}$ measured at 30~K.
Compared with the 1.5 K data, the most obvious changes in the $\sigma_{x,y,z}^{\rm SF}$ data
are the suppression of the resonance and the low-energy spin gap. Fig.~3(f) plots the energy dependence of the extracted, Bose-factor divided $M(110)$ and $M(001)$ at 30 K. In addition to the disappearance of the 
low-temperature spin gap, it can be seen that $\chi^{\prime\prime}_{110}$ still has a broad peak near $E=7$ meV, while
$\chi^{\prime\prime}_{001}$ is again relatively featureless. Comparison of the Figs.~3(e) and 3(f) reveals clear evidence for the resonance peak at $7$ meV above a spin gap of $\sim$3 meV
in $\chi^{\prime\prime}_{110}$, while $\chi^{\prime\prime}_{001}$ is featureless near 7 meV with a spin gap 
of $E\leq 2$ meV \cite{chi,slli}.
Previous unpolarized neutron scattering measurements found a spin gap value of about 3 meV at $\mathbf{Q}=(0.5,0.5,1)$ \cite{slli}.
Our polarized data are consistent with this as well as the 
unpolarized results \cite{zhao} on the same sample 
if we combine the extracted $M(110)$ and $M(001)$ results (See Appendix \ref{Appendix_unpolResults}).

To further understand the anisotropy of the spin fluctuations, we carried out constant-energy
scans with all three $\sigma_{x,y,z}^{\rm SF}$ components at $E=2.5$, 7, and 11 meV [Figs.~4(a)-(c)].
At 2.5~meV, below the spin gap, there is a peak at the in-plane wavevector $(0.5,0.5)$ in $\sigma_{z}^{\rm SF}\sim M(001)$,
whereas for the identical scan $\sigma_{y}^{\rm SF}\propto M(110)$ is featureless.
At low $\mathbf{Q}$ ($H\le0.4$) at 2.5~meV, 
the scattering for $\sigma_{x}^{\rm SF}$ and $\sigma_{y,z}^{\rm SF}$ have different backgrounds (see Appendix \ref{Appendix_background}). This problem is not present in the energy scans or other $Q$-scans taken, where the backgrounds $b_1$ and $b_2$ must be independent of polarization direction.
These constant-$E$ scans are consistent with the
constant-$Q$ scans in Fig.~3(e), suggesting that the spin gap in $M(110)$ is larger than that in $M(001)$.
At 7~meV, there are peaks in both channels at $(0.5,0.5)$, but the anisotropy appears to reverse, implying higher intensity in the in-plane $M(110)$ direction. Similar data are also found for $Q$-scans at 11 meV [Fig.~4(c)], consistent with the
constant-$Q$ data in Fig.~3.
Finally, we plot in Fig.~4(d) the
$L$-dependence of the $\sigma_{x,y}^{\rm SF}$ scattering at 7 meV and 30 K.
Instead of simply falling off as the Fe$^{2+}$ magnetic form factor \cite{ylee,ratcliff}, $\sigma_{x}^{\rm SF}$
peaks near $L=1$ and decreases rapidly with increasing $L$ above the non-magnetic background.  These results suggest
that the resonance in BaFe$_{1.9}$Ni$_{0.1}$As$_2$
has $c$-axis modulations similar to underdoped BaFe$_{2-x}$Ni$_{x}$As$_2$ \cite{mywang}
and is not entirely two-dimensional as in BaFe$_{1.84}$Co$_{0.16}$As$_2$ \cite{lumsden}.

\begin{figure}
\begin{center}
\includegraphics[width=0.9\linewidth]{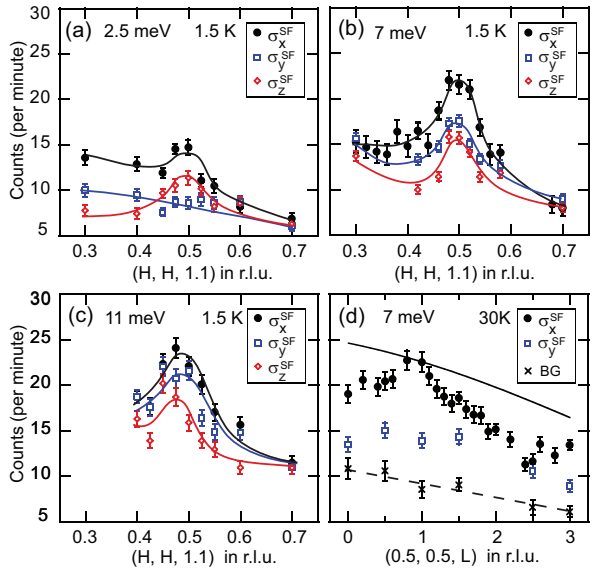}
\end{center}\caption{(color online) (a)-(c) $Q$-Scans along the $(H,H,1.1)$ direction at 2.5, 7, and 11~meV respectively, with all three spin flip cross-sections measured. (d) $L$-scan at the resonance energy.  Crosses resemble estimated instrumental background points, extracted from the data shown and $\sigma_{x,y}^{\rm NSF}$ (not shown) using Eq.~1 (assuming $b_1\approx b_2$). The solid line shows the expected magnetic scattering assuming an
Fe$^{2+}$ form factor.}
\label{Fig:fig4}
\end{figure}

\section{Conclusion}
We have performed inelastic neutron measurements with full neutron polarization analysis to measure the magnetic anisotropy of the spin fluctuations in optimally doped superconducting BaFe$_{1.9}$Ni$_{0.1}$As$_2$. We have observed the magnetic response of the iron spins pointing along in-plane [parallel to $(1,1,0)$] and out of plane [parallel to $(0,0,1)$] directions to have very different energy dependence. For the in-plane response the resonance peak was present, whereas the out of plane response was reasonably featureless around the resonance energy at 7~meV.

Spin-space anisotropy in the zero energy limit has previously been reported from NMR data on an underdoped hole-doped composition with no magnetic order \cite{matano}, which can be explained in terms of the proximity of the composition to the ordered parent compound. However, in our non-magnetically ordered sample, we see not just low frequency anisotropy, but a high frequency novel response that has different energy dependencies between different spin directions.

The presence of spin-orbital/lattice coupling could explain anisotropy 
in the spin excitations.  In pnictides, this is reflected in the undoped compound, where the 
moments are locked to the orthorhombic $a$-axis \cite{huang,zhaoprb,goldman} [along $(1,1,0)$ direction in our tetragonal notation].
The existence of the resonance solely in the in-plane response is a major challenge to the standard theory where the resonance 
is an isotropic triplet excitation of the singlet superconducting ground state.
To understand the origin of our results within the context of this theory \cite{so5}, we note
that the spin operators $\hat{S}_{110}$ and $\hat{S}_{1\bar{1}0}$, when acting on the spin singlet superconducting
ground state, generate the $S_{001}=\pm1$ components of the triplet, whereas the
spin operator $\hat{S}_{001}$ generates the $S_{001}=0$ component.  Our results therefore
imply that the resonance is the $S_{001}=\pm1$ doublet.  To understand this microscopically,
we note that in the magnetically ordered phase, the $S_{110}=0$ component of the triplet
would mix with the singlet ground state (since the moments are oriented along
the orthorhombic $a$ axis).  In the non-magnetic tetragonal state, this would lead to a low
energy doublet $S_{110}=0$, $S_{1\bar{1}0}=0$, which is equivalent to $S_{001}=\pm1$ (see Appendix \ref{Appendix_theory}).
An alternate possibility is that the resonance is instead a magnon-like excitation that becomes
undamped because of the opening of the superconducting gap \cite{morr,chubukov},
though it is not clear to us why this scenario would generate a magnetic response that is localized at
a particular energy.

The work at UT/ORNL is
supported by the U.S. DOE BES No.~DE-FG02-05ER46202, and by the U.S. DOE, Division of Scientific
User Facilities.  Work at IOP is supported by the CAS. Work at ANL is
supported by the US DOE under Contract No.~DE-AC02-06CH11357. OJL and TE were supported by the DOE BES EPSCoR Grant DE-FG02-08ER46528.

\appendix

\section{Neutron Polarization Independent Backgrounds and the 2.5~meV $Q$-cut}\label{Appendix_background}
As implied by Eq.~1 of the paper, in principle the background scattering into the detector should be the same with neutron polarization in  $x$,  $y$,  $z$ for any given SF (or NSF) measurements since the axes of the instrument do not move. However, there is a moving part that does change with neutron polarization direction, and that is the `dipole magnet' in the outgoing beam, which rotates around the scattered beam axis (with a position depending on polarization direction, as well as $\mathbf{Q}$ and $E$) and creates the neutron guide field that defines the neutron polarization direction. The problem occurs when a choice of $\mathbf{Q}$ and $E$ conspires to cause both a scattering angle that is small, and a dipole magnet position close to the horizontal for a certain neutron polarization direction. Neutrons can then scatter in grazing incidence from the dipole magnet shielding, which can increase the background in the detector for that configuration over other neutron polarization directions.

At low $\mathbf{Q}$ ($H\le0.4$) at $E=$2.5~meV, these problematic conditions are satisfied creating an extra background in the  $\sigma_{x}^{\rm NSF,SF}$ configurations.  However, the dipole magnet is away from horizontal at the same energy and wavevector for the  $\sigma_{y}^{\rm NSF,SF}$ and  $\sigma_{z}^{\rm NSF,SF}$ configurations, which therefore have lower backgrounds. At higher energies (and other wavevectors), the dipole magnet is never close to horizontal when the scattering angle is small enough for a grazing incidence to reach the detector, and so there is no difference between backgrounds of different neutron polarization configurations. We have confirmed that this is indeed the case, by comparing backgrounds extracted for all the data collected, and found an anomalous effect only for the low $\mathbf{Q}$ region at 2.5~meV.

In conclusion, at $H\le0.4$ in the 2.5~meV Q-scan there may be a difference between backgrounds in configurations with different neutron polarizations (and thus, in this case the assumption in Eq.~1 may not be valid). However, this is not a problem in any other scans, and most importantly does not affect the energy scans at any point. Therefore, as required to correctly extract $M(110)$ and $M(001)$, the assumption that the background is neutron polarization direction independent is a good one for the energy scans.

\begin{figure}
\begin{center}
\includegraphics[width=0.6\linewidth]{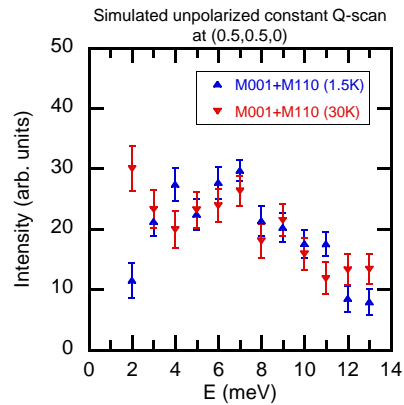}
\end{center}\caption{(color online) Intensity expected for energy scans at $(0.5,0.5,0)$ for an unpolarized experiment, calculated using $M(110)$ and $M(001)$ from the present polarized data.}
\label{Fig:fig5}
\end{figure}

\section{Comparison of Extracted Data and Previous Unpolarized Results}\label{Appendix_unpolResults}
From the present study, from observing the two different spin gaps at 3~meV and $\le2$~meV, and different maxima at approximately 7 and 3~meV in the $M(110)$ and $M(001)$ channels, one might expect to see these feautures in unpolarized data. The same compound has been previously studied \cite{zhao} by unpolarized neutrons in the $(H,K,0)$ scattering plane (different from the scattering plane in the present study).
Although there is a resonance at 7~meV and a spin gap around 3~meV,
the dynamic susceptibility does not have a peak near 3 meV.  Here we show that these results are entirely consistent with the present polarized neutron scattering results.

In the unpolarized experiment, the magnetic scattering measured at $(0.5,0.5,0)$ is proportional to $M(110)+M(001)$ for the crystal alignment used. If we assume minimal $L$ dispersion, then we can take the $M(110)$ and $M(001)$ values from our present study (where $L=1$) and simulate the ($L=0$) unpolarized data with no unknown parameters. We can then compare our simulation with the experimental data from the unpolarized experiment. As can be seen from Fig.~\ref{Fig:fig5}, the low energy features in $M(001)$ near 3 meV do not cause low energy features in the total unpolarized intensity $M(110)+M(001)$. The resulting form of Fig.~\ref{Fig:fig5} is consistent with the data in unpolarized measurements (in Ref. \onlinecite{zhao}), though the resolution in the unpolarized experiment was much better, leading to a much sharper resonance in that study.

\section{Origin of the doublet resonance}\label{Appendix_theory}
The spin singlet Cooper pair wavefunction is a product of states of the form
\begin{equation}
|\Psi_k>=|k\uparrow,-k\downarrow> - |k\downarrow,-k\uparrow>
\end{equation}We operate on this state with the spin operator, ${\hat S}$, which is the sum of $\hat{S}_1$ and $\hat{S}_2$ 
where 1 and 2 denote the two electrons of the pair. For the spin raising operator, we find\begin{equation}{\hat S}_+(q)|\Psi_k> = |k\uparrow,-k+q\uparrow> - |k+q\uparrow,-k\uparrow>\end{equation}This is the $S_z$=1 component of a triplet pair with center of mass momentum $q$ (the minus 
sign being a reflection of fermion antisymmetry).  Similarly, $\hat{S}_-$ generates the $S_z$=-1 component.
Had we operated with $\hat{S}_z$ instead, we would have obtained the $S_z$=0 component of the triplet.
Therefore, for a quantization axis along $c$, $\chi_{aa}$ and $\chi_{bb}$ generate the $S_c$=$\pm$1 
doublet, whereas $\chi_{cc}$ generates the $S_c$=0 state.  Since we find no resonance response for
$\chi_{cc}$, the resonance is the  $S_c$=$\pm$1 doublet. To better appreciate this result, assume that superconductivity and antiferromagnetism coexist, corresponding
to the spin resonance being at zero energy. If one pairs electrons using antiferromagnetic eigenstates, 
and then rewrites these pairs in terms of paramagnetic eigenstates, the resulting 
pair state is well known to be a mixture of a singlet and the $S_z$=0 component of a triplet,\cite{fenton}
with $z$ parallel to the direction of the Neel vector.  In the isotropic case, the Neel vector can point 
in any direction, which is why the resonance is a triplet.  But
for the antiferromagnetic ground state of the pnictides, the spins are locked 
to the orthorhombic $a$ axis.  Therefore, the mixed triplet component of the pairs for a coexisting state would 
be $S_a$=0.  If we then average in the plane so as to restore tetragonal symmetry, then the $S_b$=0 component 
would be involved as well.  Thus we obtain a doublet.  If we now rotate the quantization axis to be along the $c$ 
direction, this doublet corresponds to $S_c$=$\pm$1.

\end{document}